\documentclass[10pt,prl,showpacs,twocolumn]{revtex4}   

\usepackage{amsmath}    
\usepackage{graphicx}   

\begin{document}
\title{``Nonrelativistic'' kinematics: Particles or waves?}
\author{Jens Madsen Houlrik}
\affiliation{Esbjerg Institute of Technology, Aalborg University Esbjerg, Niels Bohrs Vej 8, DK-6700 Esbjerg, Denmark, European Union}
\email{jmh@aaue.dk}
\author{Germain Rousseaux}
\affiliation{Universit\'e de Nice Sophia-Antipolis, Laboratoire J.-A. Dieudonn\'e, UMR 6621 CNRS-UNS, Parc Valrose, 06108 Nice Cedex 02, France, European Union}
\email{Germain.Rousseaux@unice.fr}

\date{\today}


\maketitle

{\bf The kinematics of particles refer to events and tangent vectors, while that of waves refer to dual gradient planes. Special relativity \cite{Feigenbaum,EL06,CGK08} applies to both objects alike. Here we show that spacetime exchange symmetry \cite{Field01} implicit in the SI-definition of length based on the universal constant $c$ has profound consequences at low velocities. Galilean physics, exact in the limit $c\to\infty$, is mirrored by a dual so-called Carrollian superluminal kinematics \cite{LL65,BLL68,RVA86} exact in the limit $c\to0$. Several new results follow. The Galilean limit explains mass conservation in Newtonian mechanics, while the dual limit is a kinematical prerequisite for wavelike tachyonic motion \cite{SL1,SL2}. As an example, the Land\'e paradox \cite{Lande,LL76} of wave-particle duality has a natural resolution within special relativity in terms of superluminal, particlelike waves. It is emphasized that internal particle energy $mc^2$ can not be ignored, while kinetic energy leads to an extended Galilei group. We also demonstrate that Maxwell's equations have magnetic and electric limits covariant under Galilean and Carrollian symmetry.}

Recent essays on the status of special relativity have stressed the importance of spacetime symmetries and the associated group properties \cite{Feigenbaum,EL06,CGK08}. Inertial symmetry, leading to the principle of relativity, may be stated in terms of the familiar Galilei transformation
\begin{equation}
  \label{GT}
  {\bf r}' = {\bf r} - {\bf v}_0 t, \quad t' = t,
\end{equation}
where ${\bf v}_0=v_0\hat{\bf v}$ is the velocity of the primed frame measured in the laboratory and $\hat{\bf v}$ is a unit vector. Translational invariance means that only spacetime intervals are meaningful, and the boost (\ref{GT}) can be viewed as a shift of origin ${\bf r}'={\bf r}-{\bf r}_0$ with time dependent displacement ${\bf r}_0={\bf v}_0t$. The universal constant $c$ appears only indirectly via the SI-definition of length, and the timelike condition $r\ll ct$ must be satisfied for (\ref{GT}) to be valid. Spacelike events $r\gg ct$ behave according to the Carroll transformation \cite{LL65,BLL68,RVA86} first introduced by Jean-Marc L\'evy-Leblond
\begin{equation}
  \label{CT}
  {\bf r}' = {\bf r}, \quad t' = t - {\bf v}_0\cdot{\bf r}/c^2,
\end{equation}
where the temporal translation $t'=t-t_0$ implies loss of global simultaneity, because the offset $t_0$ depends on position. Coordinates normal to $\hat{\bf v}$ are invariant, while space-time exchange \cite{Field01} along the direction of motion
\begin{equation}
  \label{duality}
  r_\parallel\leftrightarrow ct  , \quad {\bf r}_\perp\leftrightarrow{\bf r}_\perp
\end{equation}
allows (\ref{CT}) to be obtained from (\ref{GT}) and vice-versa, in short ${\rm G}\leftrightarrow{\rm C}$. In practice, signals transmitted from a reference clock located at the origin at time $t=0$ synchronize remote clocks according to the time of travel. Using the Galilei transformation, time $t'=r'/c$ given as
\begin{equation}
  r'^2/c^2 = r^2/c^2 + \beta_0^2t^2 - 2{\bf v}_0\cdot{\bf r}t/c^2,
\end{equation}
can be approximated as
\begin{equation}
  \label{Ctime}
  t' \sim t(1+\frac{1}{2}\beta_0^2) - {\bf v}_0\cdot{\bf r}/c^2,
\end{equation}
where $\beta_0=v_0/c\ll 1$. Because coordinate transformations are linear, this result also holds for events that are not lightlike. For $r_\parallel\ll ct$, absolute time will be correct to first order in $\beta_0$. For $r_\parallel\gg ct$, the $t_0$ term must be retained, while the Galilean shift can be neglected producing (\ref{CT}). Both results are compatible with Lorentz covariance at low velocity, while the $\beta_0^2$-term in (\ref{Ctime}) represent proper time effects. Duality (\ref{duality}) implies the relations $v_\parallel\leftrightarrow V_\parallel$ and $v_\perp \leftrightarrow c\tan\alpha$, where $v_\parallel V_\parallel=c^2$ and $\tan\alpha=v_\perp/v_\parallel$ specifies direction. Low-velocity Galilean motion thus has a dual Carrollian kinematics which is superluminal \cite{SL1,SL2}. Lightlike motion is self-dual. The Lorentz transformation itself can be written in terms of dual parameters $v_0V_0=c^2$
\begin{equation}
  \label{LT}
  r_\parallel' = \gamma_0(r_\parallel - v_0 t), \quad {\bf r}_\perp' = {\bf r}_\perp, \quad
  t' = \gamma_0(t - r_\parallel/V_0),
\end{equation}
where $\gamma_0=1/\sqrt{1-\beta_0^2}$ and $\beta_0^2=v_0/V_0$. Superluminal frame velocities introduce complex numbers, because signaling will no longer be possible \cite{SLframe}.

{\it Infinitesimal transformation}.---The first-order transformation of four-events $({\bf r},ct)$ can be written in matrix form as $I-\beta L$, where the identity $I$ and the generator $L=G+C$ are $4\times 4$ matrices. The Galilean part $G$ has the components of $\hat{\bf v}$ as spatial entries in the fourth column, zero otherwise, while the Carrollian part is its transpose $C=G^T$. The full Lorentz transformation can be obtained by iteration, 
while G and C subgroups may be contracted from the Poincar\'e group \cite{CGK08,LL65,BLL68,RVA86} in the singular limit $\beta\to0$. The vector $({\bf r},t)$ transforms as
\begin{equation}
  \label{L1Event}
  I - v_0G - V_0^{-1}C,
\end{equation}
where Galilean symmetry \cite{HB03} requires $c,V_0\to\infty$, while Carrollian symmetry is obtained for $c,v_0\to0$. The respective velocity transformations show that $c'=c$ is invariant in both limits as required for a relativistic theory, and matrix expansions of the form $e^{-aX}=I-aX$, where $a$ is the parameter are exact to first order, because $G^2=C^2=0$.

{\it Particlelike versus wavelike four-vectors}.---A worldline has tangent four-velocity $\gamma({\bf v},c)$, where $\gamma=dt/d\tau$ appears as a result of proper time derivation. The vector $\gamma({\bf v},1)$ behaves according to (\ref{L1Event}). The Galilean shift
\begin{equation}
  \label{Gvel}
  \gamma'{\bf v}' = \gamma{\bf v} - {\bf v}_0 \gamma,  \quad  \gamma' = \gamma,
\end{equation}
follows for $v\ll c$, Carrollian velocity composition
\begin{equation}
  \label{Cvel}
  \gamma'{\bf v}'=\gamma{\bf v}, \quad \gamma'=\gamma(1-{\bf v}_0\cdot{\bf v}/c^2),
\end{equation}
is valid for $v\gg c$, while ${\bf v}_\perp'={\bf v}_\perp$ in both cases. Since $|\gamma(1/\beta)|=\beta \gamma(\beta)$ and $\gamma(\beta)\sim1+\frac{1}{2}\beta^2$ for $\beta=v/c\ll1$, the Lorentz factor vanishes as $|\gamma|\sim1/\beta$ at the dual speed $\beta=V/c=c/v\gg1$. Field-kinematics, on the other hand, starts from the four-gradient $(\nabla, -\partial_t/c)$. As a dual mathematical object (one-form), it has dimension of inverse length like the four-wave vector $({\bf k},\omega/c)$. Because time-components are proportional to $1/c$, the first-order transformation of $({\bf k},\omega)$ now takes the form
\begin{equation}
  \label{L1Wave}
  I - V_0^{-1}G - v_0C,
\end{equation}
where coefficients have been interchanged compared to (\ref{L1Event}). Similar considerations apply to four-momentum $({\bf p},E/c)$ and electromagnetic four-potential $({\bf A},\phi/c)$. Timelike or quasi-uniform $ck\ll\omega$ waves propagate at superluminal phase velocity given as ${\bf k}\cdot{\bf v}_p=\omega$. An order-of-magnitude estimate \cite{Houlrik09} shows that such waves are particle-like
\begin{equation}
  \label{Cwave}
  {\bf k}' = {\bf k} - {\bf v}_0 \omega/c^2,  \quad  \omega' = \omega ,
\end{equation}
in the sense that the Doppler effect is absent and wave vectors are subject to aberration due to relative time. Spacelike or quasi-stationary components $ck\gg\omega$
\begin{equation}
  \label{Gwave}
  {\bf k}' = {\bf k},  \quad  \omega' = \omega-{\bf v}_0\cdot{\bf k},
\end{equation}
show Doppler effect without aberration, because the concept of equal-time planes is frame independent in absolute time. The group velocity, given as $d{\bf k}\cdot{\bf v}_g=d\omega$, follows from dispersion $k^2-n^2\omega^2/c^2=0$, where $n$ is the refractive index of the medium. 

At first sight, particle kinematics and wave kinematics appears to be unrelated. Four-vectors deriving from events (\ref{L1Event}) are timelike for $v\ll c$, but spacelike behavior appears for $v\gg c$. On the other hand, four-vectors of the type (\ref{L1Wave}) are spacelike in the Galilean limit and timelike in the Carrollian limit. Timelike particles and spacelike waves constitute subluminal Galilean $c\to\infty$ physics. Spacelike events and timelike wave vectors are less familiar objects appearing in superluminal Carrollian $c\to0$ kinematics. In this sense, the concept of ``wave-particle duality'' originally introduced to encompass an apparent dichotomy \cite{quantons1,quantons2} takes on a new meaning in classical special relativity.

{\it Newtonian mechanics}.---The relativistic expression for mechanical four-momentum
\begin{equation}
  ({\bf p},E) = m\gamma({\bf v},c^2), 
\end{equation}
relates a wavelike vector to a particlelike vector as revealed by the factor $c^2$. The dispersion relation ${\bf p}=E{\bf v}/c^2$ can be written $vV=c^2$, where $V=E/p_\parallel$ is the energy velocity. Timelike Carrollian momentum of the form (\ref{Cwave}) equivalent to (\ref{Gvel})
\begin{equation}
  \label{Pmomentum}
  {\bf p}'={\bf p}-{\bf v}_0 E/c^2, \quad E' = E,
\end{equation}
follows for $cp\ll E$ or $v\ll c$. Newtonian momentum ${\bf p}=m{\bf v}$ is intimately related to zero-order internal energy $E=mc^2$ allowing energy transport to be replaced by mass transport with dual velocity $v$. There is no energy dispersion, since $dE/dp=0$. Galilean symmetry implies mass conservation, since mass-energy conversion becomes prohibitive for $mc^2\to\infty$. 
In contrast, spacelike Galilean momentum of the form (\ref{Gwave}) equivalent to (\ref{Cvel})
\begin{equation}
  \label{Wmomentum}
  {\bf p}'={\bf p}, \quad E' = E -{\bf v}_0\cdot{\bf p},
\end{equation}
requires $v\gg c$, a region inaccessible to ordinary particles because of the light barrier. Instead, relation (\ref{Wmomentum}) occurs frequently in areas described by Galilean wave physics \cite{Genergy} such as fluid dynamics.
We emphasize that Carrollian tachyonic ``particles'' will be wavelike objects. Using $|\gamma|\sim (c/v)(1+c^2/2v^2)$, the value $|p|=mc$ will be a minimum value, while $|E|=mc^3/v=|p|V$ decreases with speed. Being particlelike, both quantities vanish in the limit $|\gamma|\to0$. 

The action follows from the Lagrangian ${\cal L}=dS/dt={\bf p}\cdot{\bf v}-{\cal H}$ as a Lorentz invariant scalar four-product
\begin{equation}
  \label{action}
  dS = {\bf p}\cdot d{\bf r}-Edt, 
\end{equation}
which is also C and G invariant. According to (\ref{Gvel}), the free-particle Lagrangian ${\cal L}=-mc^2/\gamma$ is G invariant, because $\gamma$ is invariant at low speed. However, second-order kinetic energy enters even for $\gamma\sim 1$, because of the $c^2$-factor. The inclusion of second-order proper time effects goes beyond G and C symmetry involving terms of the form $\frac{1}{2}\beta^2(GC+CG)$. 
The second-order Lagrangian is semi-invariant ${\cal L}'={\cal L}-dS_0/dt$, where $S_0={\bf p}_0\cdot{\bf r}-E_0t$ is the center of mass action with ${\bf p}_0=m{\bf v}_0$ and $E_0= \frac{1}{2}mv_0^2$, while
\begin{equation}
  \label{Nmomentum}
  {\bf p}'={\bf p}-{\bf p}_0, \quad  E' = E - {\bf v}_0\cdot{\bf p} + E_0,
\end{equation}
reduces to the wavelike form (\ref{Wmomentum}) in the massless limit. Dispersion $E=mc^2+p^2/2m$ with $dE/dp=v$ leads to an extended Galilei group \cite{Extension}.

{\it Quantum wave-particle duality}.---According to Einstein and de Broglie \cite{BroglieN,Broglie} particles and waves $\psi\sim e^{i\varphi}$ are dual, because phase is proportional to action $S=\hbar\varphi$, where $\hbar$ is Planck's constant. Using (\ref{action}), the relations
\begin{equation}
  \label{deBroglie}
  ({\bf p},E) = \hbar({\bf k},\omega),
\end{equation}
follow, where both sides are wavelike behaving according to (\ref{L1Wave}). Particle dispersion ${\bf k}=\omega{\bf v}/c^2$ implies the dualty relation ${\bf v}\cdot{\bf v}_p=c^2$. Land\'e noticed \cite{Lande} that the spacelike Galilean transformation of de Broglie momentum $\hbar{\bf k}$ given by (\ref{Gwave}) differs from that of timelike mechanical momentum $m{\bf v}$. This long-standing paradox is resolved by space-time duality. A corpuscular Galilean particle has superluminal phase velocity and its undulatory behavior is therefore described by nonlocal Carroll kinematics (\ref{Cwave}). The corresponding momentum (\ref{Pmomentum}) does indeed become Newtonian when $E=mc^2$ is inserted. Galilean de Broglie components are less likely to appear. Fluctuations \cite{Broglie} lead to a wave packet, and the relation $p^2=E^2/c^2-m^2c^2$ is equivalent to wave dispersion with index $n^2=1-\omega_p^2/\omega^2$, where $\hbar\omega_p=mc^2$. As is well known, the relation $c^2(d{\bf k}/d\omega)\cdot({\bf k}/\omega)=n^2+n\omega(dn/d\omega)$ connects reciprocal group and phase velocities. In this case, ${\bf v}\cdot d{\bf k}/d\omega=1$ showing that ${\bf v}_g={\bf v}$.


{\it Quantum mechanics}.---Relativistic energy-momentum dispersion leads to the Klein-Gordon equation. The Schr\"odinger equation is based on $E=p^2/2m$, and the term $S_0$ therefore enters as a phase factor $\psi'= e^{-iS_0/\hbar}\psi$. One usually argue that only the norm of the wave function has physical significance \cite{LL76} allowing G, C, and L symmetry to be broken. Proper time and ``twin-paradox'' effects have been identified as missing ingredients in the Galilei transformation by Greenberger \cite{Greenberger}.

{\it ``Nonrelativistic'' electrodynamics}.---The above symmetry considerations are also relevant for electrodynamics. Using Fourier transforms the fields are defined as
\begin{equation}
  \label{EBdef}
  {\bf E} = -i{\bf k}\phi + i\omega{\bf A}, \quad {\bf B} = i{\bf k}\times{\bf A},
\end{equation}
where $({\bf A},\phi)$ is electromagnetic momentum and energy per unit charge. Faraday's law ${\bf k}\times{\bf E}=\omega{\bf B}$ and quasi-stationary $ck\gg\omega$ wave components implies the magnetic approximation $E\ll cB$ \cite{LBLL73}. From (\ref{EBdef}) it follows that the potential is spacelike $cA\gg\phi$ \cite{AJP07}. With $k\phi$ and $\omega A$ of the same order, (\ref{Gwave}) leads to the Lorentz force
\begin{equation}
  \label{GlimitEB}
  {\bf E}' = {\bf E} + {\bf v}_0\times{\bf B}, \quad {\bf B}' = {\bf B},
\end{equation}
using also $\phi'=\phi-{\bf v}_0\cdot{\bf A}$ and ${\bf A}'={\bf A}$ as well as the identity ${\bf a}\times{\bf b}\times{\bf c}=({\bf a}\cdot{\bf c}){\bf b}-({\bf a}\cdot{\bf b}){\bf c}$. The vacuum excitation fields, the displacement ${\bf D}=\epsilon_0{\bf E}$ and the field ${\bf H}={\bf B}/\mu_0$, are given in terms of permittivity and permeability related as $\epsilon_0\mu_0=1/c^2$. The transformation
\begin{equation}
  \label{ClimitDH}
  {\bf D}' = \epsilon_0{\bf E}' = {\bf D} + {\bf v}_0\times{\bf H}/c^2, \quad {\bf H}' = {\bf H},
\end{equation}
valid for $cD\ll H$ is identical to (\ref{GlimitEB}) except that the dual speed $V_0$ has replaced $v_0$. The magnetic approximation can be summarized as $E\ll Z_0H$, where $Z_0=c\mu_0=1/c\epsilon_0=\sqrt{\mu_0/\epsilon_0}$ is the vacuum impedance. The limit of infinite wave impedance $cB\to\infty$ and $cD\to0$ is Galilean for vanishing displacement $\epsilon_0\to0$ at finite $\mu_0$ and Carrollian for infinite induction $\mu_0\to\infty$ at finite $\epsilon_0$.
The electric approximation $E\gg cB$ is obtained using quasi-uniform $ck\ll\omega$ wave components and timelike potential $cA\ll\phi$. The transformation
\begin{equation}
  \label{ClimitEB}
  {\bf E}' = {\bf E},  \quad  {\bf B}' = {\bf B} - {\bf v}_0\times{\bf E}/c^2,
\end{equation}
follows using ${\bf A}'={\bf A}-{\bf v}_0\phi/c^2$ and $\phi'=\phi$. In this case,
\begin{equation}
  \label{GlimitDH}
  {\bf D}' = {\bf D},  \quad  {\bf H}' = {\bf B}'/\mu_0 = {\bf H} - {\bf v}_0\times{\bf D},
\end{equation}
valid for $cD\gg H$. The limit of zero impedance $cB\to0$ and $cD\to\infty$ is Carrollian for $\epsilon_0\to\infty$ at finite $\mu_0$ and polarization $v_0D=E/(\mu_0V_0)$ and Galilean \cite{LBLL73,AJP07} for $\mu_0\to0$ at finite $\epsilon_0$. 


Faraday's law in a medium can be written $c{\bf k}'\times c{\bf D}'=n^2\omega'{\bf H}'$, where ${\bf D}'=\epsilon{\bf E}'$ and ${\bf H}'={\bf B}'/\mu$ are defined in the medium rest frame and $\epsilon\mu= n^2/c^2$. Magnetic excitation fields $cD'\ll n^2H'$ lead to Minkowski relations \cite{EPL08}
\begin{equation}
  \label{GMin}
  {\bf D} = \epsilon{\bf E} + \epsilon(1-1/n^2){\bf v}_0\times{\bf B}, \quad  {\bf B} = \mu{\bf H},
\end{equation}
in the limit $Z_0\to\infty$, while the electric limit $cB'\to0$ and $cD'\to\infty$ lead to the Minkowski relations
\begin{equation}
  \label{CMin}
  {\bf D} = \epsilon{\bf E}, \quad {\bf B} = \mu{\bf H} - (n^2-1){\bf v}_0\times{\bf E}/c^2,
\end{equation}
in the limit $Z_0\to0$.

The macroscopic Maxwell equations with sources $\rho_s$ and ${\bf j}_s$ can be written
\begin{equation}
  \label{MaxwellEB}
  \nabla\cdot{\bf B} = 0,  \quad  \nabla\times{\bf E} = - \partial_t{\bf B},
\end{equation}
\begin{equation}
  \label{MaxwellDH}
  \nabla\cdot{\bf D} = \rho_s, \quad \nabla\times{\bf H} = {\bf j}_s + \partial_t{\bf D}.
\end{equation}
Since (\ref{MaxwellEB}) is covariant under G and C symmetry, this property is shared by (\ref{MaxwellDH}) which has the same structure in source-free regions. The symmetry of the excitation fields therefore follows from duality transformations ${\bf D}\leftrightarrow {\bf B}$ and ${\bf H}\leftrightarrow -{\bf E}$. Sources may be timelike $j_s\ll c\rho_s$ dominated by electric charge
\begin{equation}
  \label{Gcur}
  {\bf j}_s' = {\bf j}_s-\rho_s{\bf v}_0, \quad \rho_s'=\rho_s,
\end{equation}
or spacelike $j_s\gg c\rho_s$ dominated by electric current
\begin{equation}
  \label{Ccur}
  {\bf j}_s' = {\bf j}_s, \quad \rho_s'=\rho_s-{\bf v}_0\cdot{\bf j}_s/c^2,
\end{equation}
as a result of charge cancellation. The Galilean approximation (\ref{Gcur}) is particlelike (\ref{Gvel}) with invariant charge density due to the absence of length contraction. In contrast, (\ref{Ccur}) is a Carrollian transformation of the form (\ref{Cvel}). If the particle expression ${\bf j}_s=\rho_s{\bf v}_s$ is applied, superluminal speeds will result.

The Lorenz \cite{EPL05} gauge can be viewed as a continuity equation for electromagnetic four-momentum $q({\bf A},\phi/c)$, where $q$ is electric charge. Using ${\bf p}=E{\bf v}/c^2$, the potentials generated by a moving charge satisfy $A_\parallel=\phi v/c^2$. While $A_\perp$ remains unspecified, the field must be stationary $\omega'=0$ in the particle rest frame. The Galilean Doppler shift $\omega={\bf k}\cdot{\bf v}$ therefore implies ${\bf k}\cdot{\bf A}=\omega\phi/c^2$.


{\it Conclusion}.---Because $c$ is very large compared to the speed of everyday phenomena, our intuition is rooted in Galilean physics. We are familiar with particle aberration and Doppler effect as explained by local kinematics in relative space and absolute time. In contrast, spatially extended, quasi-uniform waves are nonlocal objects requiring nonlocal kinematics. Obviously, events do not have to take place in different galaxies to satisfy the condition $r\gg ct$, rather the very definition of wave fronts as equal time events implies superluminal velocities. Wave aberration without Doppler effect in relative time and absolute space is a less intuitive, particle-like phenomenon explained by nonlocal Carroll kinematics. 

As the quotation marks in the title of this Letter suggest, nonrelativistic physics reaches far beyond the simple-minded requirement of low relative frame speed. The first-order Lorentz transformation is self-dual under space-time exchange, but the Galilei transformation has the Carroll transformation as its dual and vice-versa. This observation releases the full kinematic range of special relativity needed for a resolution of the Land\'e paradox and shows the fundamental role played by space-time duality.

\end{document}